\documentclass[a4paper]{jpconf}
\usepackage{graphicx}
\begin{document}
\title{Updates of the nuclear equation of state for core-collapse supernovae and neutron Stars: effects of 3-body forces, QCD, and magnetic fields}

\author{G J Mathews$^1$, M Meixner$^1$, J P Olson$^1$, I-S Suh$^1$, T Kajino$^2$, 
T Maruyama$^{3}$, J Hidaka$^2$, C-Y Ryu$^4$, M-K  Cheoun$^5$, N Q Lan$^6$
}

\address{$^1$Department of Physics/Center for Astrophysics, University of Notre Dame, Notre Dame, United States}
\address{$^2$Theory Division, National Astronomical Observatory of Japan, Mitaka, Tokyo, Japan}
\address{$^3$College of Bioresource Sciences, Nihon University, Fujisawa 252-8510, Japan}
\address{$^4$General Education Curriculum Center, Hanyang University, Seoul 133-791, Korea}
\address{$^5$Department of Physics, Soongsil University, Seoul, South Korea}
\address{$^6$Hanoi National University of Education, 136 Xuan Thuy, Hanoi, Vietnam}

\ead{gmathews@nd.edu}

\begin{abstract}
We summarize several new developments in the nuclear equation of state for supernova simulations and neutron stars.  We discuss an updated and improved Notre-Dame-Livermore Equation of State (NDL EoS) for use in supernovae simulations.  This  Eos contains  many updates. Among them are the effects of 3- body nuclear forces at high densities and  the possible  transition to a QCD chiral and/or super-conducting color phase at densities. We  also  consider the neutron star equation of state and neutrino transport in the presence of strong magnetic fields.  We study a new quantum hadrodynamic (QHD) equation of state for neutron stars (with and without hyperons) in the presence of strong magnetic fields. The parameters are constrained by deduced masses and radii. The calculated adiabatic index for these magnetized neutron stars exhibit rapid changes with density. This may provide a mechanism for star-quakes and flares in magnetars. We also investigate the strong magnetic field effects on the moments of inertia and spin down of neutron stars. The change of the moment of inertia associated with emitted magnetic flares is shown to match well with observed glitches in some magnetars. We also discuss a perturbative calculation of neutrino scattering and absorption in hot and dense hyperonic neutron-star matter in the presence of a strong magnetic field. The absorption cross-sections show a remarkable angular dependence in that the neutrino absorption strength is reduced in a direction parallel to the magnetic field and enhanced in the opposite direction.  The pulsar kick velocities associated with this asymmetry  comparable to observed pulsar velocities and may affect the early spin down rate of proto-neutron star magnetars with a toroidal field configuration. 
\end{abstract}

\section{Introduction}
To describe the structure and hydrodynamics of compact matter;  an equation of state (EoS) is needed to relate the physics of the state variables \cite{Lattimer12}.  In supernovae the EoS determines the dynamics of the collapse and the outgoing shock, and determines whether the remnant ends up as a neutron star or a black hole.  In a neutron star, it determines the mass-radius relationship, stellar composition, cool-down time and dynamics of neutron star spin down and mergers.  In  this paper we summarize some progress in the development of Equations of state for supernova and neutron star simulations.  In particular we highlight the role of 3-body forces, QCD,  magnetic fields and neutrino transport.

\section{Three body forces and the  EoS}
At present, only a few hadronic EoSs are commonly employed that cover large enough ranges in density, temperature and electron fraction to be of use in core-collapse supernova simulations.    The two most employed in astrophysical simulations are the EoS of Lattimer \& Swesty (LS91)~\cite{LS91} and that of H.~Shen~et.~al.~(Shen98)~\cite{Shen98a, Shen98b}.  The former utilizes a non relativistic parameterization of nuclear interactions in which nuclei are treated as a compressible liquid drop including surface effects. The latter is based upon a Relativistic Mean Field (RMF) model using the TM1 parameter set in which nuclei are calculated in a Thomas-Fermi approximation. Baryonic matter was parameterized with a new RMF model that treated nuclei and non-uniform matter with the statistical model of Hempel~et.~al.~\cite{Hempel10}.

Here, we discuss  a new Notre Dame-Livermore (NDL) EoS \cite{Meixner13}.  This EoS evolves from the original Livermore formulation~\cite{Bowers82,WilsonMathews}, but unlike the previous version this NDL~EoS, is consistent with known experimental nuclear matter constraints and recent \cite{Demorest} mass and radii measurements of neutron stars. Below nuclear matter density, the conditions for nuclear statistical equilibrium (NSE) are achieved at a temperature of  $T \approx 0.5$ MeV. Below this temperature the nuclear matter is a approximated by a nine element reaction network which must be evolved dynamically. Above this temperature, the nuclear constituents are represented by free nucleons, alphas and a single ``representative'' heavy nucleus. Among the new features in the NDL EoS is that  high density phase of the EoS is treated with a parameterized Skyrme energy density functional that utilizes a modified zero range 3-body interaction.  The effects of pions on the state variables at high densities is also included as well as the consequences of a phase transition to a QGP. 

Above nuclear matter  saturation density we include both 2-body ($v_{ij}^{(2)}$) and 3-body ($v_{ijk}^{(3)}$) interactions in the many-nucleon system.  The Skyrme   two-body potential is given in the standard form~\cite{Vautherin}.  
\begin{eqnarray}
v_{12}^{(2)} &= t_0\left(1+x_0\hat{P_s}\right)\delta\left({\bf r_1 - r_2}\right) + \frac{1}{2}t_1\left(\delta\left({\bf r_1 - r_2}\right)\hat{k}^2+\hat{k}^{'2}\delta\left({\bf r_1 - r_2}\right)\right)  \\
                    &+ t_2{\bf \hat{k}^2}\cdot\delta\left({\bf r_1 - r_2}\right){\bf \hat{k}} + iW_0\left({\bf \hat{\sigma}_1}+{\bf \hat{\sigma}_2}\right)\cdot{\bf \hat{k}'} \times \delta\left({\bf r_1 - r_2}\right){\bf \hat{k}}
\label{eqn:2body}
\end{eqnarray}
Here \cite{Meixner13} we consider the possibility that the Skyrme potential  can be dominated by a 3-body repulsive interaction at high density. This term is taken to be a zero range force of the form $v_{123} = t_3\delta \left({\bf r_1 - r_2}\right)\delta \left({\bf r_2 - r_3}\right)$.  If the assumption is made that the neutron-star medium is spin-saturated~\cite{Ring}, the three-body term becomes a density dependent two-body interaction \cite{Vautherin}
that we generalize  to a modified Skyrme interaction that replaces the linear dependence on the density with a power-law index $\sigma$. \cite{Mansour}
\begin{equation}
v_{12}^{(3)'} = \frac{1}{6}t_3\left(1+\hat{P_{s}}\right)\delta \left({\bf r_1 - r_2}\right)\rho^\sigma \left(\frac{{\bf r_1+r_2}}{2}\right).
\label{eqn:3body}
\end{equation}
 A value of $\sigma$ = 1/3 is a common choice~\cite{Kohler, Krivine}.  However, in the present approach we  treat $\sigma$ as a free parameter constrained  by the skewness coefficient and observed neutron-star properties~\cite{Demorest}.

All quantities and coefficients for symmetric nuclear matter are obtained from the usual relations.
The pressure, is   $P = n^2 \frac{\partial}{\partial n} \left( \frac{E}{A} \right)$. The volume compressibility of symmetric nuclear matter is calculated from the derivative of the pressure:  $K = 9 \left(\frac{\partial P}{\partial n}\right) $  $= 18\frac{P}{n}+9n^2 \frac{\partial^2}{\partial n^2} \left( \frac{E}{A}\right)$.  The skewness coefficient, is from  the third derivative of the free energy per nucleon~$Q_0 = 27n^3\frac{\partial^3}{\partial n^3} \left(\frac{E}{A} \right) $.
 Applying the saturation condition  $P(n=n_0) = n^2 \frac{\partial}{\partial n}\left(\frac{E}{A}\right) |_{n=n_0} = 0$, one obtains \cite{Meixner13} a system of four equations in terms of $t_0$, $\left(3t_1+5t_2\right)$, $t_3$, and $\sigma$. 
Solving  
this system for $\sigma$ yields
\begin{equation}
\sigma = \frac{\frac{9}{5}T_{F_0}-2K_0 + Q_0 - 45E_0}{3K_0 + 45E_0 - \frac{27}{5}T_{F_{0}}}  ~,
\end{equation}
where the subscript zero denotes values at the saturation density.

For our purposes,  we adopt inferred values of $n_0, E_0, K_0, Q_0$ from the literature and use these to determine the Skyrme model parameters. We also demand that these parameters allow neutron star  masses $\ge 1.97 \pm 0.04 ~M_\odot$ \cite{Demorest}.  The saturation density  $n_0 \approx 0.16 ~{\rm fm}^{-3}$ and the binding energy per nucleon $E_0 = -16$ MeV are reasonably well established~\cite{BALi}. The determination of the compressibility parameter from experimental data on the giant monopole resonance on finite nuclei has been a long standing conundrum. 
For our purposes we adopt the median value and uncertainty from Ref.~\cite{Colo}, i.e.  $K_0$ = 240 $\pm$ 10 MeV as this is appropriate for the Skyrme force approach employed here. Solving  the saturation conditions  self consistently, we therefore determine the best range for the nuclear compressibility consistent with the results of~\cite{Colo}.  

There is even more uncertainty in the skewness parameter $Q_0$.  Breathing mode data ~\cite{Farine} implies  $Q_0 = -700 \pm 500$ MeV.  Using the range for $K_0$ given by~\cite{Colo} and solving 
 the saturation conditions, we find \cite{Meixner13} a skewness coefficient of  $Q_0 = -390 \pm 90$ MeV  consistent within the range given in Ref.~\cite{Colo}. The fiducial NDL~EoS is then constructed \cite{Meixner13} using the The Skyrme coefficients $t_0 =  -1718 {\rm ~ MeV fm}^{3}$, $\left(3t_1+5t_2\right) =  -102 {\rm ~ MeV fm}^5 $, $t_3 =  13226 {\rm ~ MeV fm}^{3\sigma+3}$, and $\sigma = 0.369 $.  

The density dependence of the symmetry energy beyond saturation is highly uncertain. For many Skyrme models the symmetry energy either saturates at high densities, or in the worst case becomes negative. This results in a negative pressure deep inside the neutron star core. We implemented \cite{Meixner13} a linearly increasing function of density.  The symmetry energy at saturation was  determined by the difference between the energy per particle for pure neutron matter and that of symmetric matter at  $T = 0$ MeV. For all relevant parameter sets  the NDL~EoS symmetry energy at saturation is $S_0 = 30.5$ MeV \cite{Meixner13}.

\section{QCD and the EoS}
For sufficiently high densities and/or temperature a transition from hadronic matter to quark-gluon plasma (QGP) can occur \cite{McLerran}.  Progress~\cite{Kronfeld} in lattice gauge theory (LGT) has shown that at high temperature and low density a deconfinement and chiral symmetry restoration occur simultaneously. In particular,  it has been found~\cite{Kronfeld} that the order parameters for deconfinement and chiral symmetry restoration changes abruptly for temperatures of  $T = 145 - 170$ MeV~\cite{Borsanyi, Bazavov1} as a smooth crossover. 

At low density the hadron phase can be approximated as a pion-nucleon gas, while the QGP phase can be approximated in a bag model as a non-interacting relativistic gas of quarks and gluons~\cite{Fuller}. The LGT results then imply  a  range for the QCD vacuum energy of  $165 \le  B^{1/4} \le  225$ MeV. Also requiring that  the maximum mass of a neutron star exceed  $1.97 \pm 0.04 ~M_\odot$~\cite{Demorest} implies a value for B$^{1/4}$  near the top end of that range.

For the description of quark matter we utilize a bag model with 2-loop corrections, and construct the EoS from a phase-space integral representation over scattering amplitudes.  We allow for the possibility of a coexistence mixed phase in a 1st order transition or a simple cross over transition. It is convenient to compute the QGP in terms of  the grand potential, $\Omega(T,V,\mu)$, where the grand potential for  the quark-gluon plasma takes the form:
\begin{equation}
\Omega = \sum_i(\Omega_{q0}^i + \Omega_{q2}^i) +  \Omega_{g0} + \Omega_{g2} + B V .
\label{eq:GrandPotential}
\end{equation}
Where $q_0$ and $g_0$ denote the $0^{\rm th}$-order bag model thermodynamic potentials for quarks and gluons, respectively, and $q_2$ and $g_2$ denote the 2-loop corrections. In most calculations sufficient accuracy is obtained by using fixed current algebra masses (e.g.  $m_u \sim m_d \sim 0$ GeV,  $m_s \sim 0.1 - 0.3$ GeV). For this work we chose the strange quark mass to be $m_s$ = 150 MeV and a bag constant $B^{1/4} = 165-220$ MeV.  The quark contribution to the thermodynamic potential is  given~\cite{McLerran} in terms of a sum of the ideal gas contribution plus a two loop correction from phase-space integrals over Feynman amplitudes~\cite{Kapusta}.

\begin{center}
\begin{figure}[hc]
  \centering
\includegraphics[width=0.9\textwidth]{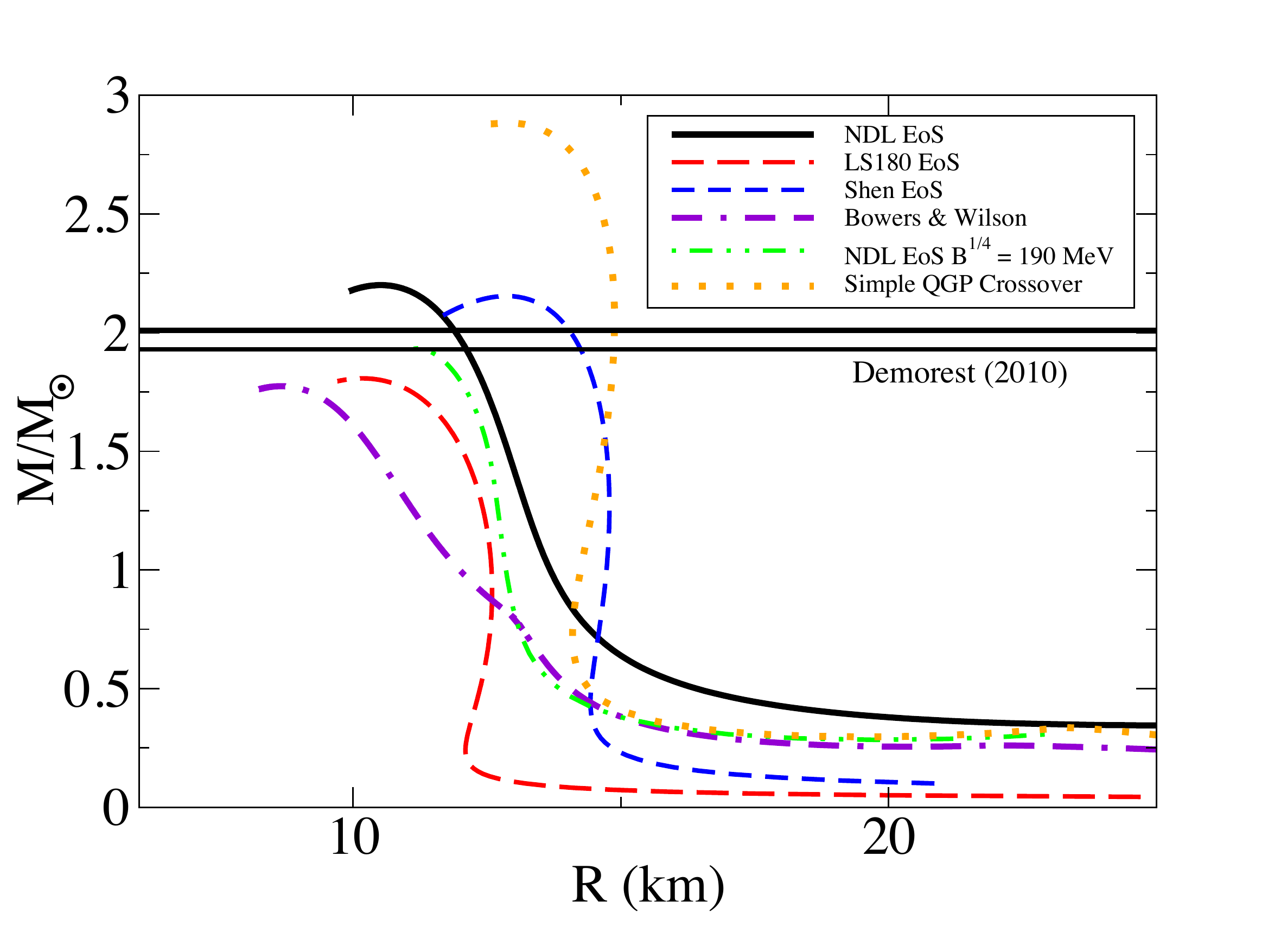}
   \caption{Mass-radius relations for the EoS of Shen \cite{Shen98a}, Lattimer~\&~Swesty (LS180) \cite{LS91}, Bowers~\&~Wilson \cite{Bowers82}, compared with the new NDL~EoS  \cite{Meixner13} with and without a mixed phase transition or a simple crossover transition to a QGP.  Note that  the soft LS180 hadronic EoS and the  previous Bowers \& Wilson EoS without 3-body forces  cannot satisfy the astrophysical constraint of a maximum neutron star mass  $\ge 1.97 \pm 0.04 ~M_\odot$ \cite{Demorest} as shown.}
 \label{fig:MaxMass}
\end{figure}
\end{center}

Fig.~\ref{fig:MaxMass} compares   \cite{Meixner13} the neutron star mass radius relation for the NDL~EoS for: 1) a hadronic EoS with 3-body forces (solid line); 2) a first order QCD transition with $B^{1/4}$ = 220 MeV; and 3) a simple QCD cross over transition. Also, shown for comparison are results from the LS180 EoS, Shen EoS and the original Bowers \& Wilson EoS. Note, that all three versions of the NDL~EoS easily accommodate a maximum neutron star  mass $\ge 1.97 \pm 0.04 ~M_\odot$, however, the hadronic version must have the  3-body forces at high baryon density. A first order phase transition to a QGP is consistent with the high maximum neutron star mass constraint~\cite{Demorest} for a bag constant  $B^{1/4} > 220$ MeV. This imposes a low baryon density transition temperature of T$_c$ = 158 MeV \cite{Fuller} which is consistent with the current range of crossover temperatures determined from LGT~\cite{Kronfeld}.

\section{Magnetic Fields and The Eos}

We have also considered  the neutron star equation of state and neutrino transport in the presence of strong magnetic fields 
\cite{Suh01}-\cite{Maruyama13}.  Indeed, magnetic fields are everywhere in Nature and frequently play a role in astrophysical phenomena.  In particular the existence of magnetars and magnetar flares \cite{duncan92,mag1,mag2}, along with the observed asymmetry in supernova explosions and the observed pulsar kick velocities all suggest the strong magnetic fields play an important role in supernova explosions and the formation of proto-neutron stars \cite{lyne94,pac92,mag3}.  In view of this we have undertaken studies of a variety of phenomena.

In \cite{Suh01} we considered the role that strong interior magnetic fields ($B \sim 10^{17}$ G) would have on neutron star structure and stability.  
We considered  the nuclear equation of
state for an ideal $npe$ gas in a strong magnetic field. In
particular, we calculated the proton concentration, the
threshold densities for neutron, muon, and pion production
and pion condensation in a strong magnetic field both
without and with the effect of the nucleon anomalous magnetic
moments. It was shown \cite{Suh01} that
the higher Landau levels are significant at high density in
spite of the existence of a very strong magnetic field. In
particular, at high density, the proton concentration
approaches the nonmagnetic limit.  In particular, we have obtained the neutron
appearance threshold  density in a magnetic field when the nucleon
anomalous magnetic moment is included. We also have shown \cite{Suh01} that the muon and pion
threshold densities are not affected by magnetic fields for  $B < 10^{17}$ G. We also  obtained an equation
of state for  a pion condensate in strong magnetic
fields. We found \cite{Suh01} that magnetic fields reduce the amount of pion condensation.
However, we still could find distinguishable effects from  a
pion condensate in strongly magnetized
neutron stars. In addition, we demonstrated  an  oscillatory behavior
of the adiabatic index in both strongly magnetized $n, p, e$
and $n, p. e, \mu, \pi$ gases at high density. Here we speculated that this
behavior might  lead to an interior pulsational instability.

In \cite{Suh10} we investigated the possibility that soft gamma-ray repeaters (SGRs) and anomalous X-ray pulsars (AXPs)
might be observational evidence for a magnetic phase separation in magnetars. We studied such magnetic domain formation
as a new mechanism for SGRs and AXPs in which magnetar matter separates into phases containing different
flux densities. We identified the parameter space in matter density and magnetic field strength at which there is an
instability for magnetic domain formation. We showed that such instabilities will likely occur in the deep outer
crust for the magnetic Baym, Pethick, and Sutherland (BPS) model and in the inner crust and core for magnetars described
in relativistic Hartree theory. Moreover, we estimated that the energy released by the onset of this instability is
comparable with the energy emitted by SGRs.

In \cite{Ryu12a} this has recently been extend to a  study a new quantum hadrodynamic (QHD) equation of state for neutron stars (with and without hyperons) in the presence of strong magnetic fields. The parameters were  constrained by the condition that deduced neutron star masses and radii that must be consistent with the recent observations \cite{Demorest} of a high mass neutron star. The calculated adiabatic index for these magnetized neutron stars exhibited the same  rapid changes with density. This was hypothesized  to provide possible  insight into the mechanism of star-quakes and flares in magnetars. We also investigated the strong magnetic field effects on the moments of inertia of neutron stars. The change of the moments of inertia associated with emitted magnetic flares was shown to match well with observed glitches in some magnetars.

In \cite{Maruyama12,Maruyama13} we explored  a perturbative calculation of neutrino scattering and absorption in hot and dense hyperonic neutron-star matter in the presence of a strong magnetic field. We found that the absorption cross-sections show a remarkable angular dependence in that the neutrino absorption strength is reduced in a direction parallel to the magnetic field and enhanced in the opposite direction.  This asymmetry in the neutrino absorption can be as much as 2 \% of the entire neutrino momentum for an large interior magnetic field.  We estimate the associated pulsar kick velocities associated with this asymmetry in a fully relativistic mean-field theory formulation and show that the kick velocities are comparable to observed pulsar velocities. In \cite{Maruyama13} we have extended this calculation to include a toroidal magnetic field configuration.  In this case, there can be an asymmetric emission of neutrino momentum along the magnetic field lines that are in the direction of the neutron star spin.  This can substantially accelerate the spin down of a neutron star in the early cooling phase, $\sim 10$ sec after core bounce. This is to be compared with \cite{Ryu12b} in which we considered the spin down of a neutron star purely from the outflow of neutrinos without a magnetic field.

\ack
Work at the University of Notre Dame is supported by the U.S. Department of Energy under Nuclear Theory Grant DE-FG02-95-ER40934. One of the authors (N.Q.L.) was supported in part by the National Science Foundation through the Joint Institute for Nuclear Theory (JINA)


\section*{References}
\begin{thebibliography}{9}
\bibitem{Lattimer12}  Lattimer J M  2012 {\it Ann. Rev. Nucl. and Part. Sci.} {\bf 62} 485

\bibitem{LS91} Lattimer J M  and  Swesty F D 1991 {\it Nuclear Physics} A {\bf  535}  331 

\bibitem{Shen98a} Shen H,  Toki H,   Oyamatsu K and Sumiyoshi K 1998  {\it Nuclear Physics} A {\bf 637} 435

\bibitem{Shen98b} Shen H,  Toki H,   Oyamatsu K and Sumiyoshi K 1998  {\it Progress of Theoretical Physics} {\bf 100} 1013 

\bibitem{Hempel10}   Hempel M and Schaner-Bielich J 2010 {\it Nuclear Physics} A {\bf 837} 210 

\bibitem{Meixner13} Meixner M,  Olson J P, Mathews G J, Lan N Q and Dalhed H E 2013  Submitted to {\it Phys. Rev.} C

\bibitem{Bowers82}Bowers R L and Wilson J R 1982 {\it Phys. Rev.} C {\bf 50} 115 

\bibitem{WilsonMathews}  Wilson J R and  Mathews G J  2003 {\it Relativistic Numerical Hydrodynamics}  (Cambridge University Press) 
\bibitem{Vautherin}  Vautherin D and Brink D M 1972  {\it Phys. Rev.} C {\bf 5} 626 
\bibitem{Ring} Ring P  and Schuck P 2000 {\it The nuclear many-body problem} (Springer)
\bibitem{Mansour} Mansour H M M 1990 {\it Acta Physica Polonica} B {\bf 21} 741 
\bibitem{Kohler} Kohler H S 1965 {\it Phys. Rev.} {\bf 138} 831 
\bibitem{Krivine} Krivine H, Treiner J and Bohigas O 1980  {\it Nuclear Physics} A {\bf 336} 155 

\bibitem{Demorest} Demorest P B, Pennucci T, Ransom S M, Roberts M S E and Hessels J W T 2010 {\it Nature} {\bf 467} 1081 

\bibitem{BALi} Li BA and Ko C M 1997  {\it Nuclear Physics} A  {\bf 618} 498  

\bibitem{Colo} Colo G, van Giai N, Meyer J, Bennaceur K and Bonche P 2004  {\it Phys. Rev.} C {\bf 70} 024307 

\bibitem{Farine} Farine M, Pearson J M and  Tondeur F 1997 {\it Nuclear Physics}  A {\bf 615} 135 

\bibitem{McLerran} McLerran L 1986 {\it Reviews of Modern Physics} {\bf 58} 1021 

\bibitem{Kronfeld}  Kronfeld A S 2012  {\it Ann. Rev. Nucl.  and Particle Sci.} {\bf 62}  265 

\bibitem{Borsanyi} Borsanyi S, et al 2012  {\it Journal of High Energy Physics} {\bf 8} 126

\bibitem{Bazavov1} Bazavov A, et al 2012 {\it Phys. Rev.} D {\bf 86} 094503  
\bibitem{Fuller}  Fuller G M,  Mathews G J and  Alcock C R 1988  {\it Phys. Rev.} D {\bf 37} 1380 
\bibitem{Kapusta} Kapusta J I 1978 {\it  High temperature matter and heavy ion collisions}  Ph.D. thesis, Univ. California , Berkeley 

\bibitem{BrSp04}Braithwaite J and Spruit H C 2004  {\it Nature} {\bf 431} 891

\bibitem{Suh01}  Suh I-S and  Mathews G J 2001 {\it Astrophys. J.}  {\bf 546}  1126  

\bibitem{Suh10}  Suh I-S and   Mathews G J 2010  {\it Astrophys. J.}  {\bf 717} 843  

\bibitem{Ryu12a} Ryu C-Y, Cheoun M-K, Maruyama T and   Mathews G J  2012 {\it Astroparticle Physics}  {\bf 38} 25-30

\bibitem{Ryu12b} Ryu C-Y, Maruyama T, Kajino T,  Mathews G J and Cheoun M K  2012 {\it Phys. Rev.} C {\bf 85}  045803 

\bibitem{Maruyama12} Maruyama T,  Yasutake N,  Cheoun M-K,  Hidaka J.  Kajino T  and Mathews  G J 2012  {\it  Phys. Rev.} D {\bf 86} 123003  

\bibitem{Maruyama13} Maruyama T,  Hidaka J.  Kajino T, Yasutake N, Cheoun M-K, Ryu C Y and Mathews  G J 2013    Submitted to {\it Phys. Rev. Lett. } ({\it Preprint} arXiv:1301.7495 [astro-ph])



\bibitem{duncan92} Duncan R C and Thompson  C 1992 {\it Astrophys. J. Lett.}  {\bf 392} L19 
\bibitem{mag1} Kouveliotou C, et al 1998 {\it Nature} {\bf 393} 235 

\bibitem{mag2} Hurley K, et al 1999 {\it Astrophys. J.} {\bf 510} L111 

\bibitem{lyne94} Lyne A G and Lorimer D R 1994 {\it Nature} {\bf 369} 127 

\bibitem{pac92} Paczy\'{n}ski B 1992  {\it Acta. Astron.} {\bf 41} 145 

\bibitem{mag3}  Chanmugam G 1992  {\it Ann. Rev. Astron. Astrophys.} {\bf 30} 143 


\end{thebibliography}
\end{document}